\documentclass[english,aps, prl,twocolumn]{revtex4-1}
\usepackage[T1]{fontenc}
\usepackage{geometry}
\usepackage{verbatim}
\usepackage{amsmath}
\usepackage{amssymb}

\makeatletter
\usepackage{graphicx}
\usepackage{xcolor}
\usepackage[bookmarks=false,linkcolor=blue,urlcolor=blue,colorlinks,citecolor=blue]{hyperref}
\usepackage[utf8]{inputenc}

\makeatother

\usepackage{babel}
\begin{document}
\title{Floquet Majorana bound states in voltage-biased Planar Josephson Junctions}
\author{Changnan Peng,$^{1,2}$ Arbel Haim,$^{1,3,4}$ Torsten Karzig,$^{5}$
Yang Peng,$^{6,1,3}$ and Gil Refael$^{1,3}$}
\affiliation{$^1$Institute for Quantum Information and Matter and Department of Physics, \mbox{California Institute of Technology, Pasadena CA 91125, USA}\\
\mbox{$^2$Department of Physics, Massachusetts Institute of Technology, Cambridge, MA 02139, USA}\\
\mbox{$^3$Walter Burke Institute for Theoretical Physics, California Institute of Technology, Pasadena, CA 91125, USA}\\
\mbox{$^4$AWS Center for Quantum Computing, Pasadena, CA 91125, USA}\\
\mbox{$^5$Microsoft Quantum, Station Q, University of California, Santa Barbara, CA 93106, USA}\\
\mbox{$^6$Department of Physics and Astronomy, California State University, Northridge, CA 91330, USA}}
\date{\today}
\begin{abstract}
We study a planar Josephson junction under an applied DC voltage bias
in the presence of an in-plane magnetic field. Upon tuning the bias
voltage across the junction, $V_{{\rm J}}$, the two ends of the junction
are shown to simultaneously host both $0$- and $\pi$-Majorana modes.
These modes can be probed using either a Scanning-Tunneling-Microscopy
measurement or through resonant Andreev tunneling from a lead coupled
to the junction. While these modes are mostly bound to the junction's
ends, they can hybridize with the bulk by absorbing or emitting photons.
We analyze this process both numerically and analytically, demonstrating
that it can become negligible under typical experimental conditions.
Transport signatures of the $0$- and $\pi$-Majorana states are shown
to be robust to moderate disorder.
\end{abstract}
\maketitle

\section{Introduction\label{par:intro}}

For some time, the idea of Floquet Majoranas has been an intriguing
concept that brought together the fields of non-abelian anyons, quantum
computing, and quantum dynamics. Floquet Majorana states were first
proposed in Ref.~\citep{Jiang2011Majorana}, and since have been
the focus of much discussion. The most direct impact that Floquet
Majoranas had was conceptual. Being excitations that are pinned to
the quasi energy which is half the drive frequency, the Floquet Majoranas
are the archetype of the time crystal phenomenon~\citep{Khemani2016phase,Else2016Floquet,vonKeyserlingk2016absolute,Yao2017discrete,Zhang2017observation}. 

The study of Majoranas in driven systems is also motivated by the
need to expand our control tools of quantum information processing
elements. A drive can also enhance the functionality of standard platforms
for non-abelian excitations. Recently, it was recognized as a way
to expand the effective dimensionality of a Majorana system, allowing
braiding of Majoranas even in a strict one-dimensional (1d) wire system~\citep{Bomantara2018simulation,Bomantara2018quantum,Bauer2019topologically}.
Also, by using the drive-induced synthetic dimensions concept~\citep{Yuan2016photonic,OzawaSynthetic2016,Martin2017topological},
one can use drives to expand the number of non-abelian anyons that
can be supported on a single Majorana wire, for instance Ref.~\citep{Peng2018time}. 

How does one best realize Floquet Majorana states experimentally?
Ref.~\citep{Liu2019floquet} proposes a realization that is in line
with the original proposal in Ref.~\citep{Jiang2011Majorana}. This
proposal involves the oscillation of the gate voltage applied to the
system to produce the time dependence needed. Other proposals include
driven quantum dots~\citep{Li2014Tunable} and quantum wires~\citep{Reynoso2013unpaired,Thakurathi2013floquet,Yates2017AditiMitra,Yates2018AditiMitra}.
In this manuscript we show that Floquet Majoranas can emerge even
in much simpler systems that are currently experimentally available.
Particularly, we analyze a spin-orbit coupled strip, put in proximity
to a superconductor in each of its sides, as considered in Refs.~\citep{Hell2017two,Pientka2017topological,Hell2017Coupling,Liu2018long,Ren2019topological,Mayer2019phase,Fornieri2019evidence,Haim2019benefits,Setiawan2019Topological,Scharf2019Tuning}.

As we show below, this system naturally gives rise to two sets of
Majorana end states when a DC bias is maintained between the two superconductors.
Indeed, through the AC Josephson effect, such a system appears driven
by the Josephson frequency $\Omega=eV_{{\rm J}}/2\hbar$. We demonstrate
the appearance of Floquet Majoranas using numerical simulations, as
well as analytical arguments. We also explore the robustness of the
Floquet Majoranas to disorder.

The rest of the paper is organized as follows. We begin in Sec.~\ref{sec:The-System}
by describing the system and explain how it can give rise to a Floquet
type of topological superconductivity which gives rise to 0- and $\pi$-Majorana
states. We then perform numerical transport simulations in Sec.~\ref{sec:numerics},
demonstrating their existence in a similar way to how they should
manifest in experiment. In Sec.~\ref{sec:Photon-assisted-coupling-of}
we analyze the effect of photon-mediated coupling between Majorana
states on opposite sides of the system. We conclude and discuss future
prospects in Sec.~\ref{sec:discussion}.

\section{The System\label{sec:The-System}}

\begin{figure}
\begin{tabular}{c c c}
\includegraphics[clip=true,trim=0mm 0mm 0mm 0mm,height=2.2cm]{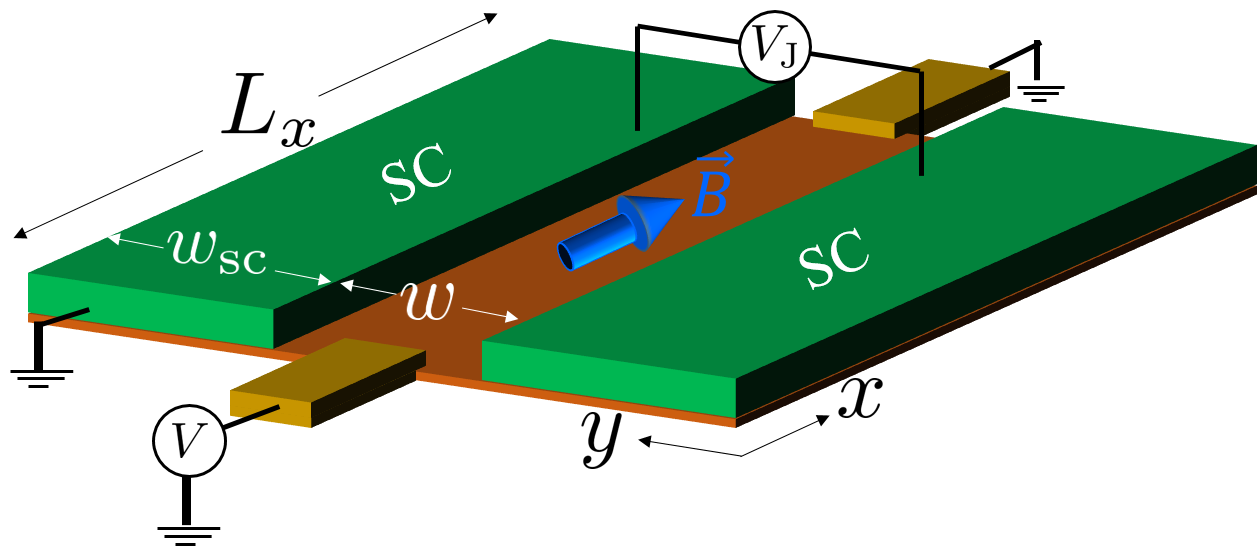}
\llap{\parbox[c]{0cm}{\vspace{-4.8cm}(a)}\hskip 5.2cm}
&
\quad
&
\hskip -4mm
\includegraphics[clip=true,trim=0mm 1.4mm 0mm 0mm,height=2.4cm]{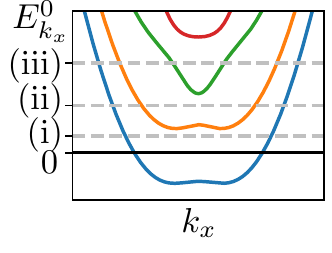}
\llap{\parbox[c]{0cm}{\vspace{-4.8cm}(b)}\hskip 4cm}
\end{tabular}

\caption{(a) Proposed experimental setup. A voltage-biased planar Josephson
junction implemented in a Rashba spin-orbit coupled 2DEG in the presence
of an in plane-magnetic field. (b) Low energy spectrum of the 2DEG
strip in the absence of superconductivity. The Fermi level ($\varepsilon=0$)
is marked by a solid black line, and $\varepsilon=\Omega/2$ is marked
by a gray dashed line, where three different scenarios are considered.
A $\pi$-Majorana state emerges at the junction ends whenever $\Omega/2$
crosses an odd number of bands. \label{fig:System}}
\end{figure}

We consider a Josephson junction, constructed by proximitizing two
conventional superconductors to a two-dimensional electron gas (2DEG)
with an in-plane applied magnetic field. Importantly, a DC bias voltage,
$V_{{\rm J}}$, is applied across the junction between the two superconductors.
To probe the system using electric transport, we further consider
two normal-metal leads coupled to the two ends of the junction. The
system is depicted in Fig.~\ref{fig:System}(a).

The Hamiltonian describing this system in the absence of the normal-metallic
leads is described by
\begin{equation}
\begin{split}\mathcal{H}(t) & =\left[-\frac{\nabla^{2}}{2m_{{\rm e}}}-\mu(y)+U(x,y)-i\alpha\left(\sigma_{y}\partial_{x}-\sigma_{x}\partial_{y}\right)\right]\tau_{z}\\
 & +E_{{\rm Z}}(y)\sigma_{x}+{\rm Re[}\Delta(y)]\tau_{x}-{\rm Im}[\Delta(y)]\tau_{y}.
\end{split}
\label{eq:H_PJJ}
\end{equation}
This Hamiltonian is written in the basis of the Nambu spinor $\Psi^{\dagger}(\boldsymbol{r})=[\psi_{\uparrow}^{\dagger}(\boldsymbol{r}),\psi_{\downarrow}^{\dagger}(\boldsymbol{r}),\psi_{\downarrow}(\boldsymbol{r}),-\psi_{\uparrow}(\boldsymbol{r})]$,
where $\psi_{s}^{\dagger}(\boldsymbol{r})$ creates an electron inside
the 2DEG at position $\boldsymbol{r}=(x,y)$ with spin $s$. The Pauli
matrices $\{\sigma_{x,y,z}$\} and $\{\tau_{x,y,z}$\} operate on
the spin and particle-hole degrees of freedom, respectively. Here,
$m_{{\rm e}}$ is the effective electron mass in the 2DEG, $\mu(y)=\mu_{{\rm J}}\Theta(w/2-|y|)+\mu_{{\rm SC}}\Theta(|y|-w/2)$
is the chemical potential, $\mu_{{\rm J}}$ ($\mu_{{\rm SC}}$) being
its value in the junction (below the superconductors), where $\Theta$
is the Heaviside step function and $y=0$ denotes the middle of the
system, $U(x,y)$ is a disorder potential due to impurities, $\alpha$
is the Rashba spin-orbit coupling coefficient, and $E_{{\rm Z}}(y)=E_{{\rm Z}}^{0}\Theta(w/2-|y|)$
is the Zeeman splitting due to the in-plane magnetic field present
in the junction. Finally, the induced superconducting potential inside
the 2DEG is given by $\Delta(y)=\Delta_{0}\Theta(|y|-w/2)\exp[i\Theta(y)\phi(t)]$,
where a linearly time-dependent phase bias $\phi(t)=2eV_{{\rm J}}t$
is generated by the voltage across the junction.

As a result of the oscillating phase between the superconductors,
the Hamiltonian of Eq.~(\ref{eq:H_PJJ}) is time periodic, $\mathcal{H}(t+T)=\mathcal{H}(t)$,
with a period $T=\pi/(eV_{{\rm J}})$. We can accordingly write the
Hamiltonian using its Floquet representation,
\begin{equation}
\mathcal{H}_{mn}^{{\rm F}}=n\Omega\delta_{mn}+\frac{1}{T}\int_{0}^{T}{\rm d}te^{-i(m-n)\Omega t}\mathcal{H}(t),\label{eq:H_F}
\end{equation}
where $\Omega=2\pi/T=2eV_{{\rm J}}$, and $m,n\in\mathbb{Z}$. By
construction, the spectrum of $\mathcal{H}^{{\rm F}}$ is periodic
under $\varepsilon\to\varepsilon+\Omega$. The Floquet Hamiltonian
further obeys a particle-hole symmetry, $\tau_{y}\sigma_{y}[\mathcal{\mathcal{H}^{{\rm F}}}_{m,n}]^{\ast}\tau_{y}\sigma_{y}=-\mathcal{\mathcal{H}^{{\rm F}}}_{-m,-n}$,
dictating a symmetry of the spectrum under $\varepsilon\to-\varepsilon$.
Together with the periodicity of the spectrum, one concludes that
a single state with either $\varepsilon=0$ or $\varepsilon=\Omega/2$
is protected and cannot be removed by any perturbation respecting
these symmetries~\citep{Jiang2011Majorana}. Such states are referred
to as 0-Majorana and $\pi$-Majorana states, respectively, where $0$
and $\pi$ correspond to the phase acquired by these states upon a
unitary evolution over a time $T$.

To gain some intuition, one can first consider the weak pairing limit.
In this limit the induced superconducting pairing inside the junction
can be treated as a small perturbation to the band structure of an
isolated semiconducting strip. This band structure, shown in Fig.~\ref{fig:System}(b),
contains multiple transverse bands which are spin-split due to spin-orbit
coupling and magnetic field. As in the stationary case of a topological
superconductor~\citep{Kitaev2001unpaired,Alicea2012,Beenakker2013,Aguado2017majorana,Lutchyn2018majorana},
one expects 0-Majorana states to emerge when the Fermi level (black
solid line), $\varepsilon=0$, crosses an odd number of bands (namely
an odd number of pairs of Fermi points). In the case of a driven (Floquet)
topological superconductor, one expects, in addition, $\pi$-Majorana
states to emerge whenever the line $\varepsilon=\Omega/2$ (gray dashed
line) crosses through an odd number of bands. Below, we consider three
different values of $V_{{\rm J}}$ corresponding to $\varepsilon=\Omega/2$
crossing either one, two, or three bands.

Throughout this work, we take the system parameters to be $\Delta_{0}=500\mu{\rm eV}$,
$E_{{\rm so}}=m_{{\rm e}}\alpha^{2}/2=100\mu{\rm eV}$, $l_{{\rm so}}=\hbar/(m_{{\rm e}}\alpha)=100{\rm nm}$,
$\mu_{{\rm J}}=37.5\mu{\rm eV}$, $\mu_{{\rm SC}}=1{\rm m}{\rm eV}$,
$E_{{\rm Z}}^{0}=75\mu{\rm eV}$, $w=292{\rm nm}$, and $w_{{\rm sc}}=292{\rm nm}$.
This corresponds to $m_{{\rm e}}=3.47\times10^{-32}{\rm kg}=0.038m_{{\rm e}}^{0}$
and $\alpha=3.04\times10^{4}{\rm m/s}$.

\section{Numerical Analysis\label{sec:numerics}}

To simulate the system numerically, we truncate the Floquet indices,
$m,n\in[-N_{{\rm F}},N_{{\rm F}}]$ in Eq.~(\ref{eq:H_F}). For a
large-enough cutoff $N_{{\rm F}}$ this is justified by the frequency-space
localization of the Floquet eigenstates, which is induced by the term
$n\Omega\delta_{mn}$. We further discretize the Hamiltonian $\mathcal{H}_{mn}^{{\rm F}}$
spatially by constructing an appropriate tight-binding Hamiltonian
on a rectangular lattice. In the present work we keep 7 Floquet bands
($N_{{\rm F}}=3$), and take the lattice constants to be $a_{x}=40{\rm nm}$,
and $a_{y}=73{\rm nm}$.

{} 

To probe the presence of Majorana modes we consider the case where
two normal-metallic leads are connected to the two ends of the junction,
as depicted in Fig.~\ref{fig:System}(a). In this setup, the presence
of $0$- and $\pi$-Majorana states at the junction ends will induce
resonant Andreev reflection of electron arriving from one of the normal
metal leads with energy $0$ and $eV_{{\rm J}}$, respectively. Experimentally,
this should be observed in the DC differential conductance, $\sigma(V)=dI_{{\rm DC}}/dV$,
where $I_{{\rm DC}}$ is the DC component of the current in the normal-metal
lead and $V$ is its voltage with respect to ground.

To obtain this quantity numerically, we calculate the scattering matrix
of the discretized truncated Floquet Hamiltonian, with the reflection
and transmission blocks having the form%
,
\begin{equation}
r_{im;jn}=\begin{pmatrix}r_{im;jn}^{{\rm ee}} & r_{im;jn}^{{\rm eh}}\\
r_{im;jn}^{{\rm he}} & r_{im;jn}^{{\rm hh}}
\end{pmatrix}\,\,\,;\,\,\,t_{im;jn}=\begin{pmatrix}t_{im;jn}^{{\rm ee}} & t_{im;jn}^{{\rm eh}}\\
t_{im;jn}^{{\rm he}} & t_{im;jn}^{{\rm hh}}
\end{pmatrix}.
\end{equation}
For example, $r_{im;jn}^{{\rm he}}\equiv r_{ij}^{{\rm he}}(\varepsilon+m\Omega,\varepsilon+n\Omega)$
is the amplitude for an electron in mode $(i,m)$ to be reflected
as a hole in mode $(j,n)$, where $m,n$ label the Floquet sectors
and $i,j$ each label the spin and transverse modes in the lead. The
scattering matrix is calculated using the recursive Green function
technique~\citep{Lee1981Anderson} (see Refs.~\citep{Haim2019benefits,Haim2019quantum}
for details of implementation).

The DC differential conductance can then be extracted from the scattering
matrix using the Landauer-B\"uttiker formalism, generalized for a
periodically-driven superconducting system,
\begin{equation}
\begin{split} & \sigma(V)=\frac{e^{2}}{h}\sum_{ij}\sum_{n=-\infty}^{\infty}\Big[|t_{ij}^{{\rm ee}}(eV,eV+n\Omega)|^{2}\\
 & +|t_{ij}^{{\rm he}}(eV,eV+n\Omega)|^{2}+2|r_{ij}^{{\rm he}}(eV,eV+n\Omega)|^{2}\Big].
\end{split}
\label{eq:dI_dV}
\end{equation}
Terms involving $t^{{\rm ee}}$ and $t^{{\rm he}}$ describe processes
where a single electron is emitted, therefore contributing a unit
quantum conductance, while the Andreev reflection term $r^{{\rm he}}$
describes a process where two electrons are emitted from the lead
and therefore contributes two units of quantum conductance~\citep{Blonder1982transition}.
Unlike the case of a stationary system, however, each of these processes
can now occur through an absorption or emission of $n$ photons~\citep{Moskalets2002floquet,Moskalets2012scattering,Kundu2013transport}.

\begin{figure*}
\begin{centering}
\begin{tabular}{c c c}
\includegraphics[clip=true,trim=0mm 0mm 0mm 0mm,height=4cm]{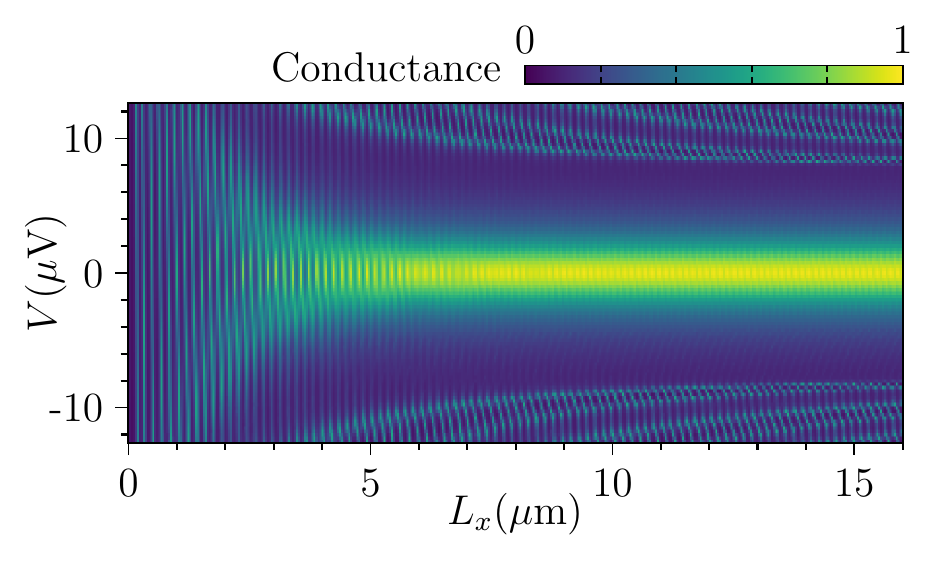}
\llap{\parbox[c]{0cm}{\vspace{-7cm}(a)}\hskip 7cm}
&
\quad\quad
&
\includegraphics[clip=true,trim=0mm 0mm 0mm 0mm,height=4cm]{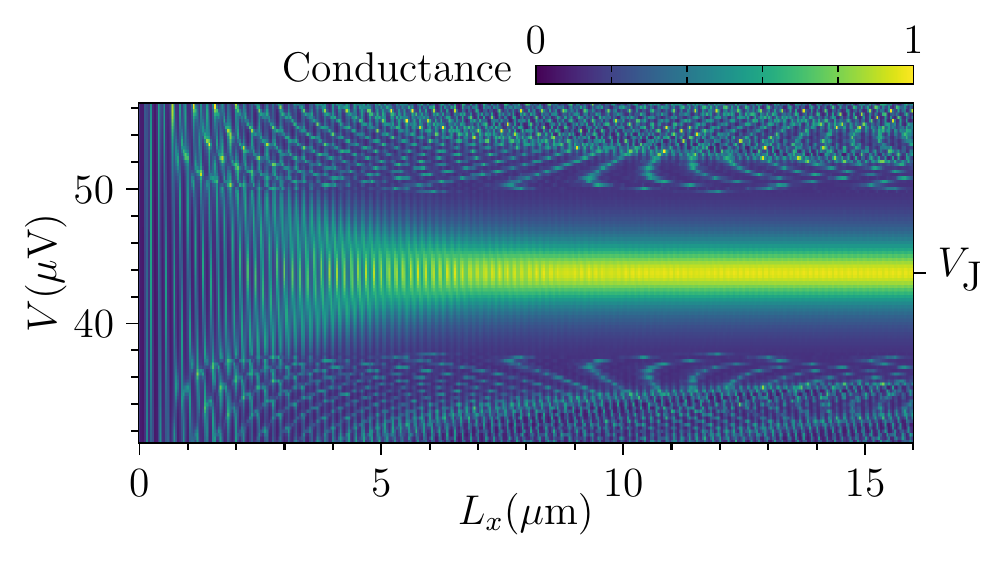}
\llap{\parbox[c]{0cm}{\vspace{-7cm}(b)}\hskip 7.5cm}
\\
\includegraphics[clip=true,trim=0mm 0mm 0mm 0mm,height=4cm]{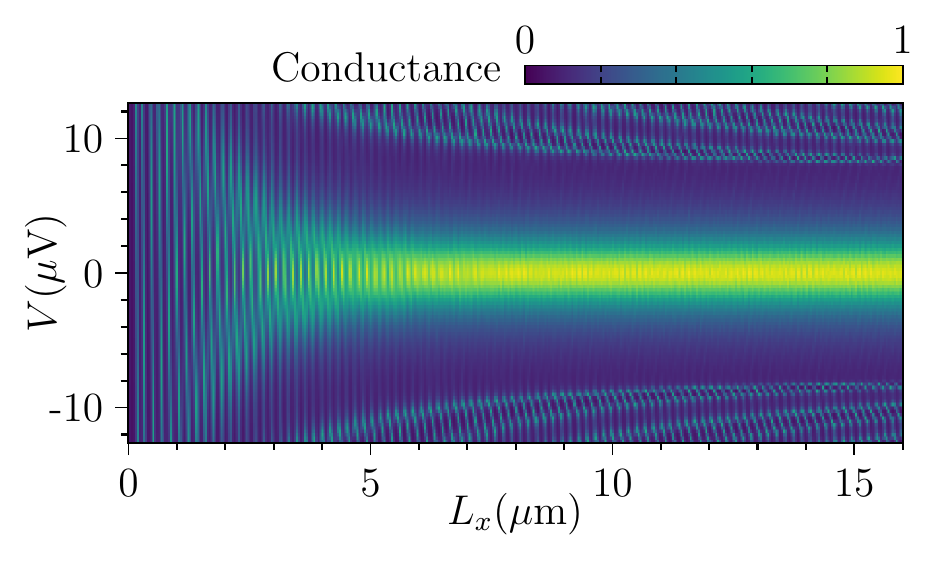}
\llap{\parbox[c]{0cm}{\vspace{-7cm}(c)}\hskip 7cm}
&
\quad\quad
&
\includegraphics[clip=true,trim=0mm 0mm 0mm 0mm,height=4cm]{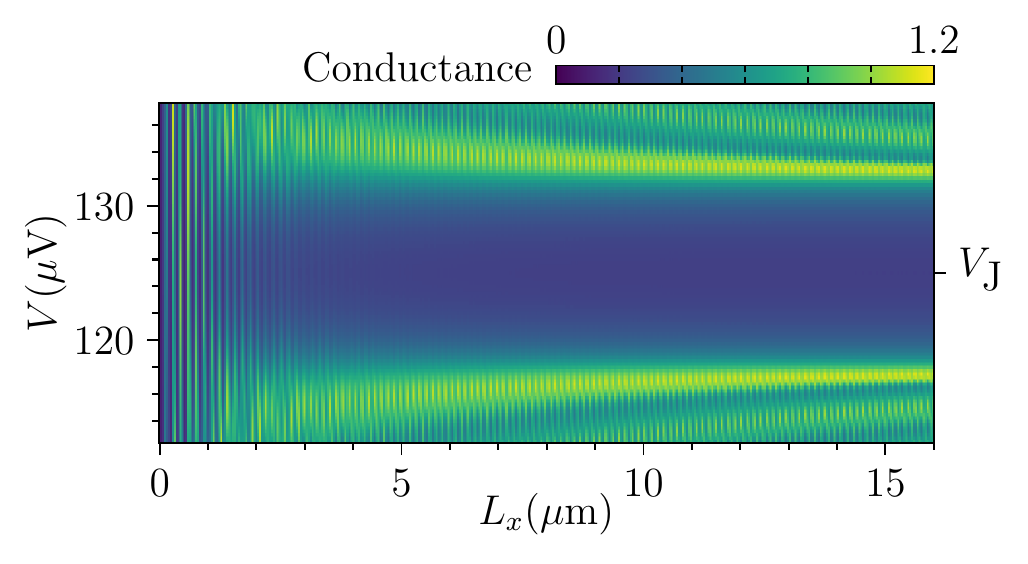}
\llap{\parbox[c]{0cm}{\vspace{-7cm}(d)}\hskip 7.5cm}
\\
\includegraphics[clip=true,trim=0mm 0mm 0mm 0mm,height=4cm]{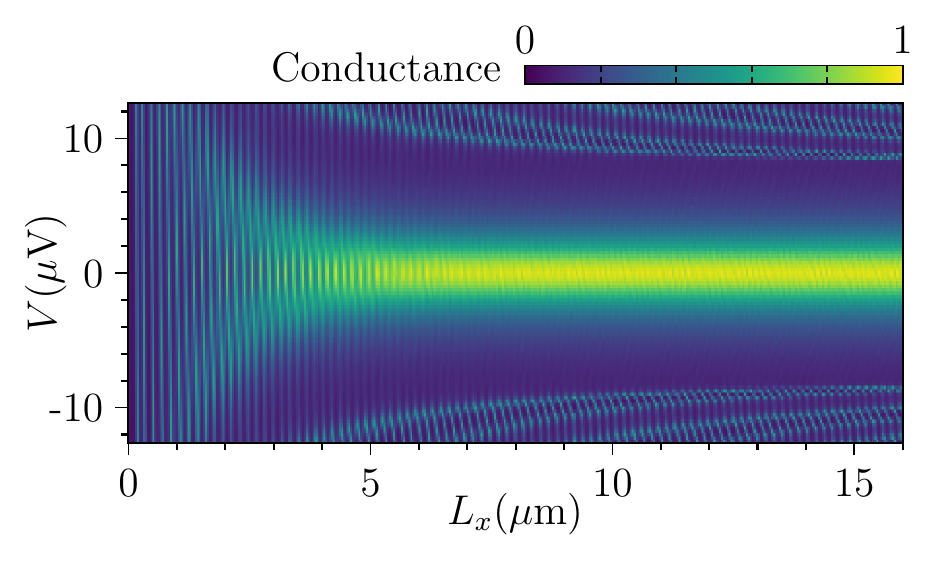}
\llap{\parbox[c]{0cm}{\vspace{-7cm}(e)}\hskip 7cm}
&
\quad\quad
&
\includegraphics[clip=true,trim=0mm 0mm 0mm 0mm,height=4cm]{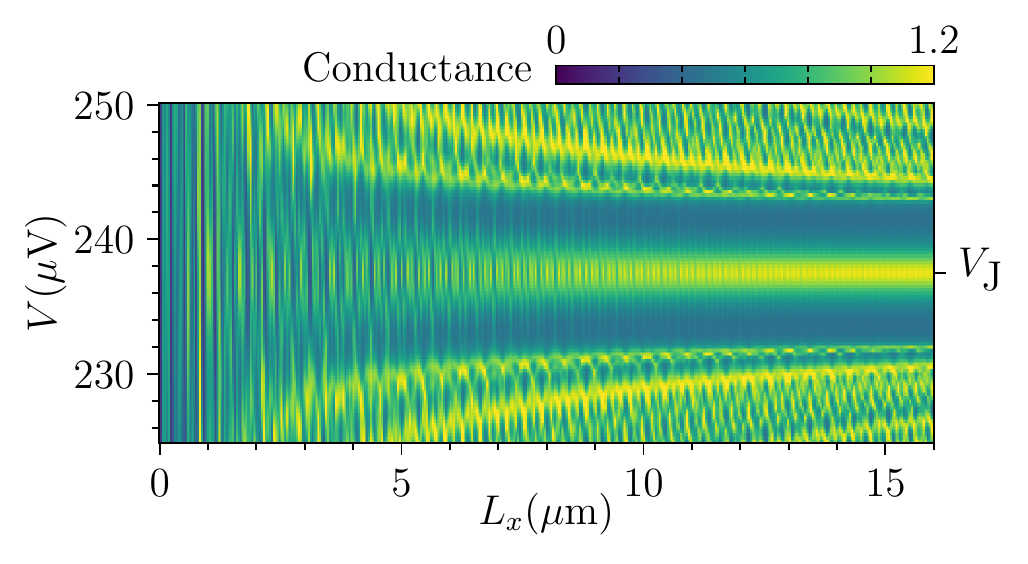}
\llap{\parbox[c]{0cm}{\vspace{-7cm}(f)}\hskip 7.5cm}
\end{tabular}
\end{centering}

\caption{Conductance as a function of the lead's voltage $V$ and the junction's
length $L_{x}$, in units of $2e^{2}/h$. The voltage across the junction
is taken to be $V_{\textrm{J}}=43.75\mu\textrm{V}$ (a,b), $V_{\textrm{J}}=125\mu\textrm{V}$
(c,d), and $V_{\textrm{J}}=237.5\mu\textrm{V}$ (e,f). These three
values correspond to the the three dashed gray lines in Fig.~\ref{fig:System}(b),
marked (i), (ii), and (iii), respectively. \label{fig:Conduct_vs_L_and_Vb}}
\end{figure*}

In Fig.~\ref{fig:Conduct_vs_L_and_Vb}, we present results for $\sigma(V)=dI_{{\rm DC}}/dV$
as a function of the junction's length $L_{x}$ {[}see Fig.~\ref{fig:System}(a){]},
and the voltage in the lead $V$, for a clean system. The left panels,
Fig.~\ref{fig:Conduct_vs_L_and_Vb}(a,c,e), focus on voltages near
$V=0$, while the right panels, Fig.~\ref{fig:Conduct_vs_L_and_Vb}(b,d,f),
focus on voltages near $V=V_{{\rm J}}$. For a long-enough system,
the emergence of $0$- and/or $\pi$-Majorana states can be seen as
a robust resonance at $V=0$ and/or $V=V_{{\rm J}}$, respectively.

The top, middle, and bottom panels correspond to three different values
of the voltage $V_{{\rm J}}$ across the junction. These three values
of $V_{{\rm J}}$ are shown in Fig.~\ref{fig:System}(b) as gray
dashed lines, labeled (i), (ii) and (iii). It is chosen such that
$\varepsilon=\Omega/2=eV_{{\rm J}}$ crosses either a single band
{[}Fig.~\ref{fig:Conduct_vs_L_and_Vb}(a,b){]}, two bands {[}Fig.~\ref{fig:Conduct_vs_L_and_Vb}(c,d){]},
or three bands {[}Fig.~\ref{fig:Conduct_vs_L_and_Vb}(e,f){]}. As
expected, $\pi$-Majorana modes emerge when the number of bands crossed
by $\varepsilon=\Omega/2$ is odd. In all these cases the chemical
potential $\mu$ is kept constant with the Fermi level ($\varepsilon=0$)
crossing a single band as shown in Fig.~\ref{fig:System}(b). Signatures
of $0$-Majorana states can accordingly be seen in all the left panels
of Fig.~\ref{fig:Conduct_vs_L_and_Vb}.

In the stationary case of a topological superconductor, under some
general conditions the conductance resonance is quantized to $\sigma(0)=2e^{2}/h$,
at zero temperature~\citep{Bolech2007Observing,Law2009majorana,Flensberg2010tunneling,Fidkowski2012universal}.
More specifically, if the system is gapped and long enough, the transmission
matrix vanishes and the reflection matrix can be shown to have a single
perfectly-Andreev-reflecting channel %
{} in the topological phase~\citep{Beenakker2011random}. If in addition
the lead is weakly coupled to the system (or if the lead has a single
channel), the rest of the channels do not contribute, yielding $\sigma(0)=2e^{2}/h.$

In contrast, for a periodically driven topological superconductor
the resonances at $V=0$ and $V=V_{{\rm J}}$ are generally not quantized,
even in the limit of weakly coupled lead and a gapped infinite system.
Instead, quantization is only obtained when summing over the differential
conductance at certain discrete energies~\citep{Kundu2013transport,Farrell2016transport}.
In the presence of a $0$-Majorana bound state one has $\sum_{m}\sigma(2meV_{{\rm J}})=2e^{2}/h$,
while in the presence of a $\pi$-Majorana bound state one has $\sum_{m}\sigma[(2m+1)eV_{{\rm J}}]=2e^{2}/h$.

The resonances seen in Fig.~\ref{fig:Conduct_vs_L_and_Vb}(a-c,e)
exhibit a peak values only slightly less than $2e^{2}/h$. This can
suggest that conductance at $V=2meV_{{\rm J}}$ and $V=(2m+1)eV_{{\rm J}}$
with $m\neq0$ are relatively suppressed. This is reasonable considering
that electrons arriving at these energies require the absorption or
emission of several photons in order be in resonance with the Majorana
states. For $|\Delta_{{\rm ind}}|<|\Omega|$, which is the case considered
here, such processes would be suppressed. In the case of Fig.~\ref{fig:Conduct_vs_L_and_Vb}(f),
on the other hand, the conductance resonance exhibit a peak value
slightly above $2e^{2}/h$. This could be a result of the coupling
to the lead being comparable with the induced gap, allowing for higher-energy
states to contribute to conductance. Indeed, the induced gap around
$V_{{\rm J}}$ shown in Fig.~\ref{fig:Conduct_vs_L_and_Vb}(f) is
smaller than the gaps seen in the spectra of Fig.~\ref{fig:Conduct_vs_L_and_Vb}(a-e).

\begin{figure}
\includegraphics[clip=true,trim=0mm 0mm 0mm 0mm,width=8.5cm]{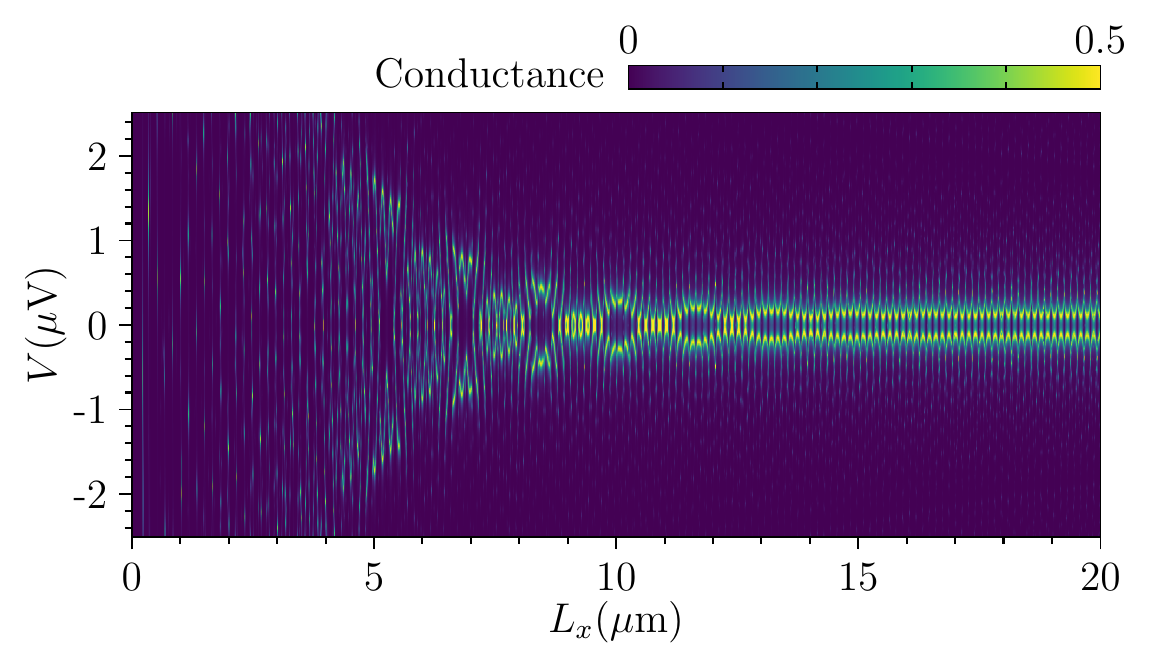}

\caption{Conductance as a function of the lead's voltage $V$ and the junction's
length $L_{x}$, in units of $2e^{2}/h$. System parameters are the
same as in Fig.~\ref{fig:Conduct_vs_L_and_Vb}(a,b), except the coupling
to the lead is reduced, enabling to view the splitting of the Majorana
resonance. The splitting does not decrease to zero with increasing
$L_{x}$, and is most likely a result of photon-mediate coupling between
Majorana states on opposite ends of the junction (see Sec.~\ref{sec:Photon-assisted-coupling-of}).
\label{fig:Conduct_high_resol}}
\end{figure}

To examine the Majorana-induced resonance with better resolution,
we consider the conductance for the case shown in Fig.~\ref{fig:Conduct_vs_L_and_Vb}(a),
but with a weaker coupling between the system and the normal-metal
leads. This causes the width of the resonance to decrease, allowing
for a closer examination of the splitting of the Majorana modes. The
result is shown in Fig.~\ref{fig:Conduct_high_resol}. One can now
clearly observe the splitting of the resonance away from $V=0$. As
the junction's length, $L_{x}$ is increased, the resonance energy
initially oscillates with a decreasing amplitude, however, beyond
about $L_{x}\sim10\mu{\rm m}$ the splitting approaches a constant
value. This behavior is quite different than that of a static topological
superconductor, where the asymptotic splitting of the Majorana modes
is exponentially decaying. The behavior observed here is most likely
due to photon-induced coupling between the Majorana states at the
two ends of the junction. In this process, a photon can excite a quasiparticle
into a conducting mode of the system, allowing for cross talk between
the Majorana end states. In Sec.~\ref{sec:Photon-assisted-coupling-of}
we analyze this splitting and show that it becomes small whenever
$|k_{F}\xi|\gg1$ or when $|\Omega|\gg|\mu|$, where $k_{{\rm F}}$
is the Fermi momentum and $\xi$ is the Majorana localization length
in the static case.

\begin{figure}
\begin{tabular}{c}
\includegraphics[clip=true,trim=0mm 0mm 0mm 0mm,width=8.5cm]{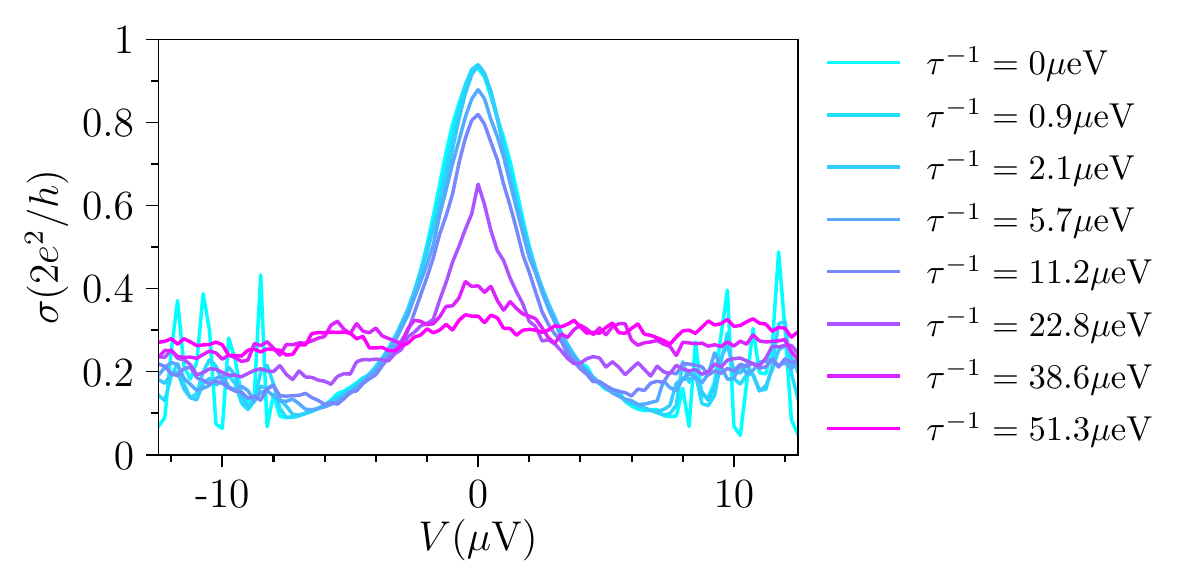}
\llap{\parbox[c]{0cm}{\vspace{-8cm}(a)}\hskip 8.5cm}
\\
\includegraphics[clip=true,trim=0mm 0mm 0mm 0mm,width=8.5cm]{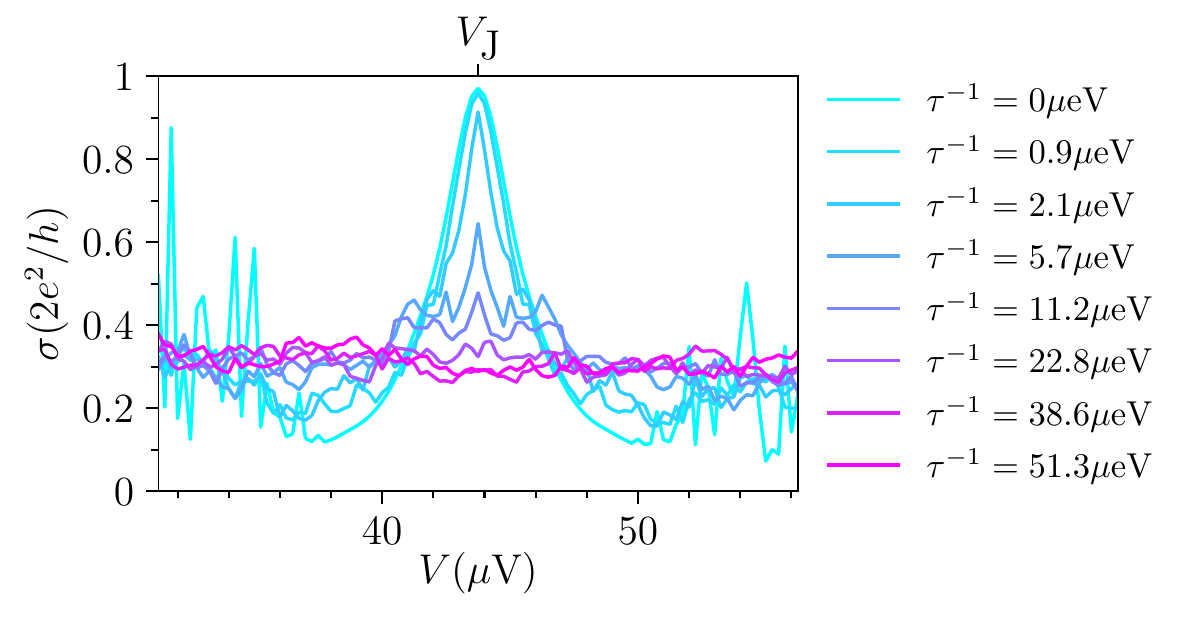}
\llap{\parbox[c]{0cm}{\vspace{-8cm}(b)}\hskip 8.5cm}
\end{tabular}

\caption{Conductance as a function of the lead's voltage $V$ for a fixed junction's
length, $L_{x}=16\mu{\rm m}$, and different values of disorder strength
represented by inverse mean-free time. System parameters are the same
as in Fig.~\ref{fig:Conduct_vs_L_and_Vb}(a,b). In (a) we focus on
voltages around $V=0$ and in (b) on voltages around $V=V_{{\rm J}}$,
corresponding to $0$- and $\pi$-Majorana states, respectively. \label{fig:Conductance_disorder}}
\end{figure}

We end the section on numerical results by demonstrating the robustness
of the signatures observed above to weak disorder. We focus on the
system parameters used in Fig.~\ref{fig:Conduct_vs_L_and_Vb}(a,b),
and simulate random short-correlated disorder, $U(\boldsymbol{r})U(\boldsymbol{r}')=\delta(\boldsymbol{r}-\boldsymbol{r}')/(m_{{\rm e}}\tau)$.
Here, $\tau$ is the disorder-induced mean-free time in the case of
unproximitized 2DEG and in the absence of a magnetic field. In Fig.~\ref{fig:Conductance_disorder},
we present results for the conductance as a function of the lead's
voltage for different values of $\tau$. Each data point is a result
of averaging over 50 disorder realizations. Figures~\ref{fig:Conductance_disorder}(a)
and \ref{fig:Conductance_disorder}(b) focus on voltages near $V=0$
and $V=V_{{\rm J}}$ corresponding to $0$- and $\pi$-Majorana states,
respectively. In both cases the Majorana-induced (nearly quantized)
peak remains intact for a finite range of disorder strengths, beyond
which the peak value begins to decrease until disappearing completely.
Notice the critical value of the disorder correspond to $\tau^{-1}$
which is of the order of the induced gap in the system, as observed
in Fig.~\ref{fig:Conduct_vs_L_and_Vb}(a,b), resembling the behavior
of the static topological superconductor in the presence of disorder~\citep{Motrunich2001Griffiths,Brouwer2011Probability,Lobos2012interplay,Pientka2013disorder,Huse2013disorder,Rieder2013reentrant,Adagideli2014disorder,Rieder2014Density}.

\section{Photon-induced coupling of the Majorana modes\label{sec:Photon-assisted-coupling-of}}

We have seen above that while the signatures of the Majorana states
are robust, there is a small splitting of the Majorana resonance that
does not decay with increasing the system size, as apparent in Fig.~\ref{fig:Conduct_high_resol}.
Such a splitting does not exist in a static topological superconductor,
and is a consequence of the periodic drive induced by the voltage
bias across the junction.

To analyze this effect we consider a simplified version of the system.
We treat the region inside the junction as a 1d semiconductor weakly
coupled to the superconductors {[}see Fig.~\ref{fig:System}(a){]}.
For adequate values of the chemical potential and magnetic field,
this system is known to be described by a spinless \emph{p}-wave superconductor~\citep{Alicea2012}.
Since in our case a voltage bias is applied between the superconductors,
the Hamiltonian for the system reads
\begin{equation}
H_{p}(t)=\sum_{k}\xi_{k}c_{k}c_{k}^{\dagger}+\frac{1}{2}[(1+e^{-i\Omega t})\Delta_{k}c_{k}^{\dagger}c_{-k}^{\dagger}+{\rm h.c.}]\label{eq:H_p(t)}
\end{equation}
where $\Omega=2eV_{{\rm J}}$ as before, $\xi_{k}$ is the dispersion
of the lowest electronic band, and $\Delta_{k}=-\Delta_{-k}$ is the
effective \emph{p}-wave pairing potential. For concreteness, we take
$\Delta(k)=\Delta'k$ and $\xi_{k}=k^{2}/2m-\bar{\mu}$ for some effective
values of the electron mass $m$, the chemical potential $\bar{\mu}$,
and the coefficient $\Delta'$. We emphasize, however, that in the
weak pairing limit, the physics is determined by $\Delta_{k}$ and
$\xi_{k}$ near the Fermi momentum $k_{{\rm F}}$, defined by $\xi_{k_{{\rm F}}}=0$.
We assume this limit below. 

In the absence of the time-dependent term, the Hamiltonian of Eq.~(\ref{eq:H_p(t)})
has a gap at the Fermi level ($\varepsilon=0$). The periodic-time-dependent
term, however, enables a process in which by absorbing a photon a
quasiparticle can be excited into a conducting mode. Focusing on the
0-Majorana, we retain only the $n=0$ Floquet sector and Floquet bands
which can be reached by absorbing at most a single photon to leading
order in $|\Delta(k_{{\rm F}})|/\Omega$. The resulting Hamiltonian
can be written in first-quantization form as
\begin{equation}
\mathcal{H}_{p}^{{\rm F}}=\frac{1+\lambda_{z}}{2}(\xi_{k}\tau_{z}+\Delta'k\tau_{x})+\frac{1-\lambda_{z}}{2}(\xi_{k}-\Omega)\tau_{z}+\Delta'k\lambda_{x}\tau_{x}.
\end{equation}
where $\lambda_{x,y,z}$ are Pauli matrices operating on the space
of states having $\{0,1\}$ photons. The first term above corresponds
to the 0-photon sector and describes a static spinless $p$-wave superconductor.
The second term corresponds to the 1-photon sector and describes a
gapless 1d channel. Finally, the third term couples the two sectors
and describes electron pairing mediated by an absorption or emission
of a photon. We note that a similarly-structured Hamiltonian can be
obtained for describing the $\pi$-Majorana by focusing on quasi-energies
near $\varepsilon=\Omega/2$ instead of $\varepsilon=0$.

To make analytic progress, we first treat the 0-photon sector by solving
for the Majorana end-states, $\gamma_{{\rm L}}$ and $\gamma_{{\rm R}}$,
that emerge in the presence of open-boundary conditions and projecting
out the rest of the spectrum. We then integrate out the 1-photon modes
to obtain a self-energy term describing a coupling between the two
Majorana end states, in addition to the exponentially-small finite-size
coupling. The result for the $2\times2$ Green function of the ground-state
manifold, written in the Nambu basis $(\gamma_{{\rm L}},\gamma_{{\rm R}})$
is given by
\begin{equation}
G^{{\rm R}}(\omega)=[\omega-\epsilon_{{\rm M}}\tau_{y}-\Sigma(\omega)]^{-1},\label{eq:Green_func}
\end{equation}
where $\Sigma(\omega)$ is the self-energy due to photon-mediated
pairing, and $\epsilon_{{\rm M}}\propto\exp(-L/\xi)$ is the exponentially-decaying
energy splitting between the Majorana states in the static case, with
the decay length given by $\xi=1/(m|\Delta'|)$. We note that in the
weak pairing limit one has $k_{{\rm F}}\xi\gg1$.

In the limit of $L\gg\xi$, one can neglect $\epsilon_{M}$. The shift
and broadening of the Majorana resonance can then be obtained from
the zero-frequency self energy, which to leading order in $1/(k_{{\rm F}}\xi)$
reads
\begin{equation}
\Sigma(0)=\frac{\Delta'k_{{\rm F}}}{(k_{{\rm F}}\xi)^{2}}\left(\frac{\bar{\mu}}{\Omega}\right)^{2}\left[\frac{32}{k_{{\rm F}}\xi}(1+\frac{\bar{\mu}}{\Omega})\tau_{y}-i8\sqrt{1+\frac{\Omega}{\bar{\mu}}}\right].\label{eq:self_E}
\end{equation}
The first term in Eq.~(\ref{eq:self_E}) gives the energy splitting
of the Majorana modes, while the second term gives its broadening
and represents the hybridization of the Majorana state with the continuum
of extended modes in the wire. Considering the self-energy for finite
values of $\omega$ results in corrections to the splitting and the
broadening, however, we have verified that these involve higher order
of $1/(k_{{\rm F}}\xi)$.

From Eq.~(\ref{eq:self_E}) it is evident that, unlike in the static
case, the splitting between the Majorana states does not decay with
the length of the system. Nevertheless, in the limit of either $k_{F}\xi\gg1$,
or $\Omega\gg\bar{\mu}$, this splitting can be much smaller than
the induced superconducting gap in the system $\Delta'k_{{\rm F}}$.
Such a situation is indeed observed in the numerical simulations of
Sec.~\ref{sec:numerics} as apparent from Fig.~\ref{fig:Conduct_high_resol}.
For short enough system lengths, $\epsilon_{{\rm M}}$ dominates over
$\Sigma(0)$ in Eq.~(\ref{eq:Green_func}) and the splitting of the
Majorana modes follows oscillations with exponentially-decaying amplitude.
For longer system size, $\Sigma(0)$ dominates over $\epsilon_{{\rm M}}$
and the splitting between the Majorana modes follows a constant value.

Note that while the broadening in Eq.~(\ref{eq:self_E}) is parametrically
larger in $1/k_{F}\xi$ than the splitting, for our parameters the
larger numerical prefactor of the splitting is sufficient to lead
to similar values for splitting and broadening. In general, there
will be other contributions to the self energy, $\Sigma(0)\rightarrow\Sigma(0)+\tilde{\Sigma}$,
that go beyond the continuum modes and the splitting is just observable
for sufficiently small $\tilde{\Sigma}$. One example, mentioned above,
is the broadening by the coupling to the transport leads. Also, interactions
and phonon-induced relaxation will contribute to $\tilde{\Sigma}$.
We note that when a decay rate $\gamma$ is included in the Green
function in Eq.~(\ref{eq:Green_func}), this contribution is $\tilde{\Sigma}=2i\gamma(\Delta'k_{F})^{2}/\Omega^{2}$,
which can be neglected relative to the broadening of Eq.~(\ref{eq:self_E})
for $\gamma\ll\Delta'k_{F}\sqrt{1+\Omega/\bar{\mu}}=\Delta_{\text{ind}}\sqrt{1+\Omega/\bar{\mu}}$.
Typical phonon relaxation times correspond to $\gamma\lesssim1\mu\text{eV}$
and can therefore be neglected.

\section{Discussion\label{sec:discussion}}

We have investigated a voltage-biased Josephson junction implemented
in a two-dimensional electron gas in the presence of an in-plane magnetic
field. We have shown that this system supports a pair of weakly-coupled
$0$-Majorana end states together with a pair of weakly coupled $\pi$-Majorana
states. The weak coupling between Majorana end states on opposite
sides of the junction is induced by photon absorption or emission
which causes the Majorana modes to hybridize with the highly-excited
conducting modes. As we show, this coupling can, nevertheless, become
exceedingly small for reasonable system parameters.

For a phase-biased Josephson junction of the type we study here, it
was previously shown that the system supports zero-energy Majorana
bound states at each end of the junction~\citep{Pientka2017topological,Hell2017Coupling}.
Such a system was subsequently studied by several experimental groups
who observed signatures of topological superconductivity~\citep{Ren2019topological,Fornieri2019evidence,Mayer2019phase}.
Our results suggest that a slight modification of the same experimental
setup can realize a Floquet topological superconductor. The presence
of $0$-Majorana and $\pi$-Majorana states in such a system can be
directly probed by measuring DC differential conductance from a metallic
lead coupled to one of the junction's ends as a function of its voltage
$V$, as depicted in Figs.~\ref{fig:System}(a). This should produce
simultaneous nearly-quantized resonances at $V=0$ and $V=V_{{\rm J}}$,
respectively, the latter being the voltage bias across the junction
{[}see Fig.~\ref{fig:Conduct_vs_L_and_Vb}{]}.

Further insight into the origin of these resonances can be gained
by considering a situation where the system is physically split into
two parts at the middle of the junction ($y=0$). Each subsystem then
consists of a superconductor in proximity to a 1d semiconductor, and
can therefore be tuned into a (static) topological superconducting
phase~\citep{Oreg2010helical,Lutchyn2010majorana}, giving rise to
a Majorana bound states at each of its ends. Since the Fermi energies
of the two subsystems differ by $V_{{\rm J}}$, one pair of Majorana
bound states resides at energy $\varepsilon=0$, while the other resides
at $\varepsilon=V_{{\rm J}}$. These result in conductance peaks at
$V=0$ and $V_{{\rm J}}$, respectively. Interestingly, these features
survive even when the two subsystems are brought together, as shown
in this work. Indeed, the voltage-biased Josephson junction allows
electrons (and holes) to gain energy through multiple Andreev reflections
and escape the gap to a conducting channel, possibly hybridizing Majorana
states at opposite ends. As shown in Secs.~\ref{sec:numerics},~\ref{sec:Photon-assisted-coupling-of},
however, in practice this hybridization is rather weak.

Applying a voltage bias to the junction in search of a Floquet topological
superconductor has the added advantage of introducing a tuning parameter
to the system, $V_{{\rm J}}$, in addition to the junction chemical
potential, $\mu$, (controlled by a gate) and the in-plane magnetic
field. In the weak-pairing limit, $0$ $(\pi)$-Majorana modes should
appear whenever $\mu$ ($V_{{\rm J}}$) crosses through an odd number
of Zeeman-split bands {[}see Fig.~\ref{fig:System}(b){]}. This was
demonstrated numerically (see Fig.~\ref{fig:Conduct_vs_L_and_Vb}),
together with the robustness of the Majorana modes to disorder (see
Figs.~\ref{fig:Conductance_disorder}).

An exciting prospects of Floquet topological superconductors is the
ability to implement braiding of Majorana modes in a strictly 1d system~\citep{Bomantara2018quantum,Bomantara2018simulation,Bauer2019topologically}.
In this scenario one takes advantage of the fact that the $0$- and
$\pi$-Majorana can be thought of as residing in separate channels.
Such a process can in principle be implemented in the system considered
here, by adding local gates to control the position of the Majorana
modes, together with an additional AC potential to couple the 0- and
$\pi$-Majorana modes in restricted regions. The photon-mediated coupling
between opposite Majorana end states discussed above will in principle
cause the braiding operation to be unprotected, as it can induce a
non-universal dynamical phase. Nevertheless, one might be able to
avoid this by performing the braiding on a time scale shorter compared
with the inverse energy splitting of the Majoranas.

\section*{Acknowledgments}

This research was supported by the Institute of Quantum Information
and Matter, an NSF Frontier center funded by the Gordon and Betty
Moore Foundation, the Packard Foundation, and the Simons foundation.
AH acknowledges support from the Walter Burke Institute for Theoretical
Physics at Caltech. The results of this work were obtained prior to
the employment of AH at the Amazon Web Services Center for Quantum
Computing. GR is also grateful for support through NSF DMR grant number
1839271. This work was performed in part at Aspen Center for Physics,
which is supported by National Science Foundation grant PHY-1607611.

\appendix

\section{Numerical results for a narrow junction}

\begin{figure}[h]
\begin{centering}
\includegraphics[clip=true,trim=0mm 0mm 0mm 0mm,width=9cm]{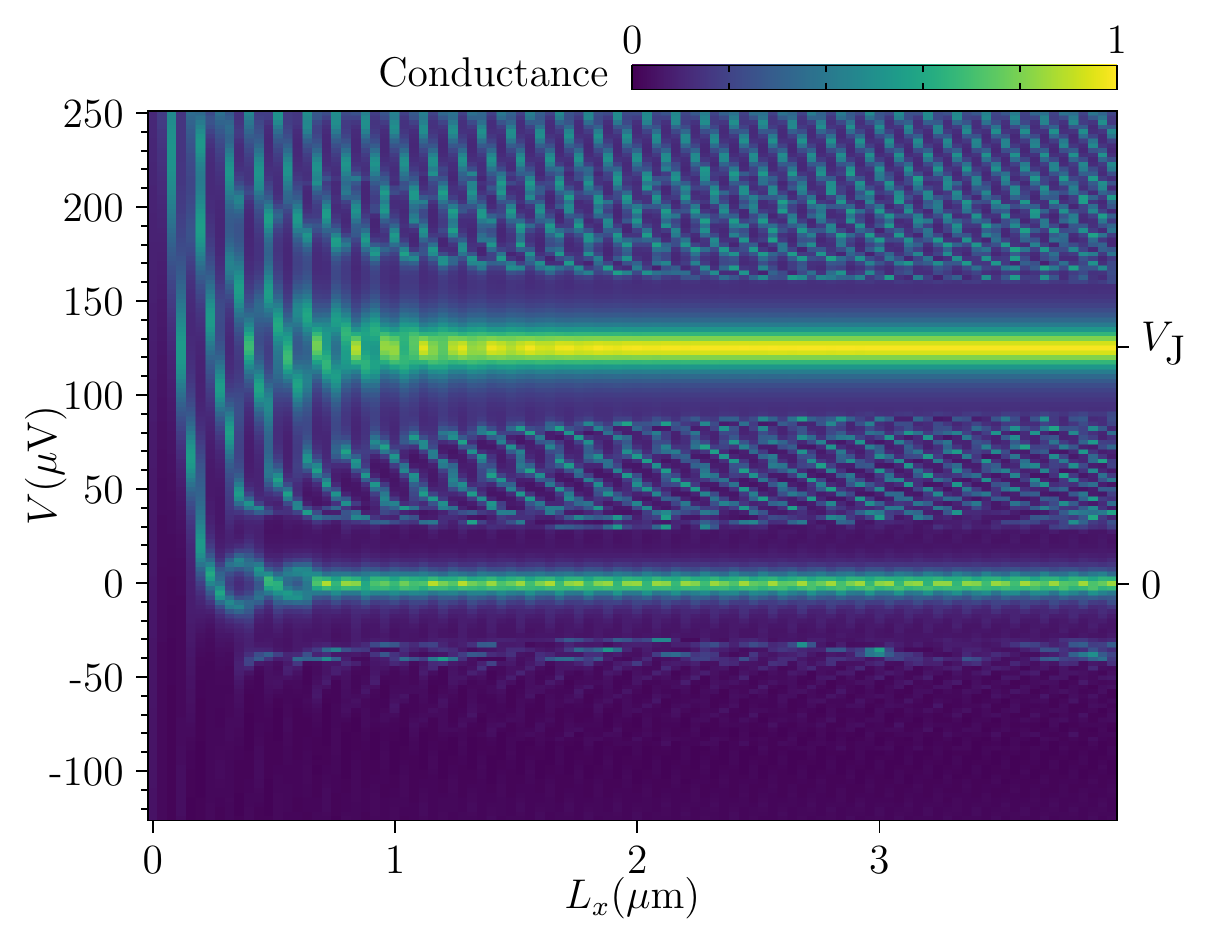}
\end{centering}

\caption{Conductance as a function of the lead's voltage $V$ and the junction's
length $L_{x}$, in units of $2e^{2}/h$. The voltage across the junction
is taken to be $V_{\textrm{J}}=125\mu\textrm{V}$.\label{fig:Conductance_narrow_junction}}
\end{figure}

Decreasing the width of the junction, $w$, is expected to increase
the induced superconducting gap in the junction. Decreasing $w$ also
causes higher transverse modes to get pushed to high energies. In
this case, the relevant scenario for observing simultaneous $0$ and
$\pi$ Majorana modes is when the lines $\varepsilon=0$ and $\varepsilon=\Omega/2=eV_{\text{J}}$
both cross a single pair of Fermi points {[}see scenario (i) in Fig.~\ref{fig:System}(b){]}.
The larger splitting between transverse modes allows one to increase
the Zeeman field and $\Omega$, thereby suppressing photon-induced
coupling of the Majorana modes to bulk conducting mode. In Fig.~\ref{fig:Conductance_narrow_junction}
we present numerical results for the differential conductance as a
function of voltage and system's length, for $w=73{\rm nm}$, $\mu_{{\rm J}}=287.5\mu{\rm eV}$,
$\mu_{{\rm SC}}=1.125{\rm m}{\rm eV}$, $E_{{\rm Z}}^{0}=250\mu{\rm eV}$,
$V_{\textrm{J}}=125\mu\textrm{V}$, keeping the rest of the system
parameters unchanged. The conductance spectrum exhibits simultaneous
resonances at $V=0$ and $V=V_{{\rm J}}$, each separated by a sizable
gap of about $40\mu{\rm eV}$.

\bibliographystyle{apsrev4-1}
\bibliography{References}

\end{document}